\documentclass[reprint,twocolumn,superscriptaddress,showpacs,nofootinbib,notitlepage,preprintnumbers,aps,prd,floatfix]{revtex4-2}

\usepackage{graphicx} 
\usepackage{amsmath}
\usepackage{amssymb}
\usepackage{mathtools}
\usepackage{natbib}
\usepackage{bbm}
\usepackage{tikz}
\usepackage{todonotes}
\usepackage{soul}
\usepackage{multirow}

\usetikzlibrary{quantikz2}
\usepackage[colorlinks=true,backref=false, linktocpage=true,
citecolor=blue,urlcolor=blue,linkcolor=blue,pdfpagemode=UseOutlines]{hyperref}
\hypersetup{%
  bookmarksnumbered=true,
  pdftitle = {},
  pdfsubject = {},
  pdfauthor = {},
  pdfkeywords = {}
}
\usepackage{xcolor}
\newcommand{\BT}{$\mathbb{BT}$}
\newcommand{\BO}{$\mathbb{BO}$}

\definecolor{some-color}{rgb}{0.75, 0.25, 0.75}

\begin{document}
\preprint{FERMILAB-PUB-24-0241-SQMS-T}

\title{Highly-efficient quantum Fourier transformations for certain nonabelian groups}

\newcommand{\GW}{Department of Physics, The George Washington University, Washington, DC 20052}

\newcommand{ \SQMS}{Superconducting and Quantum Materials System Center (SQMS), Batavia, Illinois, 60510, USA}

\newcommand{\USRA}{Universities Space Research Association, Research Institute for Advanced Computer Science (RIACS) at NASA Ames Research Center, Moffet Field, CA, 94035, USA}

\newcommand{\FNAL}{Fermi National Accelerator Laboratory, Batavia, Illinois,  60510, USA}

\author{Edison M. Murairi}
\email{emm712@gwu.edu}
\affiliation{\GW}

\author{M. Sohaib Alam}
\affiliation{Superconducting and Quantum Materials System Center (SQMS), Batavia, Illinois, 60510, USA}
\affiliation{Quantum Artificial Intelligence Laboratory (QuAIL),
NASA Ames Research Center, Moffett Field, CA, 94035, USA}
\affiliation{USRA Research Institute for Advanced Computer Science (RIACS), Mountain View, CA, 94043, USA}
\author{Henry Lamm}
\affiliation{Superconducting and Quantum Materials System Center (SQMS), Batavia, Illinois, 60510, USA}
\affiliation{Fermi National Accelerator Laboratory, Batavia,  Illinois, 60510, USA}

\author{Stuart Hadfield}
\affiliation{Superconducting and Quantum Materials System Center (SQMS), Batavia, Illinois, 60510, USA}
\affiliation{Quantum Artificial Intelligence Laboratory (QuAIL),
NASA Ames Research Center, Moffett Field, CA, 94035, USA}
\affiliation{USRA Research Institute for Advanced Computer Science (RIACS), Mountain View, CA, 94043, USA}

\author{Erik Gustafson}
\email{egustafson@usra.edu}
\affiliation{Superconducting and Quantum Materials System Center (SQMS), Batavia, Illinois, 60510, USA}
\affiliation{Quantum Artificial Intelligence Laboratory (QuAIL),
NASA Ames Research Center, Moffett Field, CA, 94035, USA}
\affiliation{USRA Research Institute for Advanced Computer Science (RIACS), Mountain View, CA, 94043, USA}

\date{\today}

\begin{abstract} 
    Quantum Fourier transformations are an essential component of many quantum algorithms, from prime factoring to quantum simulation. 
While the standard abelian QFT is well-studied, important variants corresponding to  \emph{nonabelian} groups of interest have seen less development.
    In particular, fast nonabelian Fourier transformations are important components for both quantum simulations of field theories as well as approaches to the nonabelian hidden subgroup problem. In this work, we present fast quantum Fourier transformations for a number of nonabelian groups of interest for high energy physics, $\mathbb{BT}$, $\mathbb{BO}$, $6\Delta(27)$, $\Delta(54)$, and $\Sigma(36\times3)$. 
   For each group, we derive explicit quantum circuits and estimate resource scaling for fault-tolerant implementations.  
    Our work shows that the development of a fast Fourier transformation can substantively reduce simulation costs by an up to three orders of magnitude for the finite groups that we have investigated.
\end{abstract}
\maketitle

\section{Introduction}

Fourier transforms have numerous applications in both classical and quantum computing.   
Classical applications range from signal analysis and compressed sensing, to coding theory, physics, and number theory 
\cite{stankovic2005fourier}.  
In terms of quantum algorithms, it forms an important subroutine for Shor's algorithm~\cite{Shor:1994jg}, the hidden subgroup~\cite{simon1997power,ivanyos2001efficient} and hidden shift problems~\cite{2002quant.ph.11140V,2003quant.ph..2112K}, phase estimation~\cite{cheung2003using}, and -- of particular interest here -- quantum simulation of high energy physics (HEP)~\cite{jordan2012quantum,PhysRevD.100.034518,Kan:2021xfc,hardy2024optimized}.

Shor's algorithm and its extensions for 
prime factoring and discrete logarithms rely on the ability of quantum computers to efficiently solve the hidden subgroup problem for finite abelian groups using the quantum Fourier transform (QFT).
A great deal of research has investigated implementing the abelian QFT on quantum hardware \cite{Mayer:2024yfm,Wright:2024yxm,Gois:2024aef,Baumer:2024elv,Pfeffer:2023yhb,2016Natur.536...63D,Mundada:2022roq,Cakmak:2023kvh,Jin:2023iqd}.
Beyond special cases, 
it is unknown if efficient QFT circuits imply an efficient solutions to the related hidden subgroup problem, a question connected to numerous deep open problems in quantum computing~\cite{childs2010quantum}. Indeed, the known nonabelian fast Fourier transforms are often more complex~\cite{Hoyer:1997qc,beals1997quantum, Pueschel:1998zzo,moore2003generic,childs2010quantum,kawano2016quantum} than their abelian counterparts and developing efficient algorithms remains an active area of research both classically and quantumly~\cite{childs2010quantum}. Furthermore, even for particular classes of groups where asymptotically efficient quantum circuit are known, detailed resource analysis of specific cases can yield dramatic improvements over the general techniques. 

The need for fast QFTs for nonabelian groups is 
particularly acute in the quantum simulations of HEP, where previous work found that the naive Fourier transform represented $\gtrsim 95\%$ of the total gate cost~\cite{alam2022primitive,Gustafson:2022xdt,Gustafson:2023kvd, gustafson2024primitive}. Important targets for HEP applications are the QFTs for crystallographic subgroups of $SU(2)$~\cite{SU2groups,hanany1999non,falbel2004fundamental} and $SU(3)$~\cite{fairbairn1964finite,Grimus:2010ak,Ludl:2011gn,ludl2012proceedings,bauer2013new}, and beyond those cases $SU(N)$~\cite{hanany2001monograph,parattu2011tribimaximal} more generally. 
As we will see below, the reduction in resource costs for HEP simulations from fast QFTs provide multiple orders of magnitude improvement. While we focus on groups of interest to HEP, we anticipate our work to have broader algorithmic applications. A recent example \cite{Bravyi:2023veg,ikenmeyer2023remark} utilized the QFT over the symmetric group~\cite{beals1997quantum,kawano2016quantum} within generalized quantum phase estimation~\cite{harrow2005applications} to derive quantum algorithms and complexity results concerning computing structural properties of the group known as Kronecker coefficients.

Mathematically, Fourier transforms are defined over a specific group -- either abelian or nonabelian, continuous or finite -- and correspond to a unitary transformation from the group algebra to a complex vector
space whose basis vectors are the matrix elements of the group's irreducible representations (irreps). For a finite group $G$, the \emph{discrete Fourier transform} (DFT) can be defined as
\begin{eqnarray}
\hat{f}(\rho) = \sqrt{\frac{d_{\rho}}{|G|}} \sum_{g \in G} f(g) \rho(g),
\label{eqn:Fourier-group}
\end{eqnarray}
where $\vert G \vert$ is the size of the group, $d_{\rho}$ is the dimensionality of the representation $\rho$, and $f$ is a function over $G$. Implemented naively, the abelian DFT over $\mathbb{Z}_n$ has a classical complexity of $O(n^2)$. Through repeated application of a subgroup-decomposition, it is possible to reduce the classical complexity to $O(n \log(n))$ with a \emph{fast Fourier transform} (FFT) using the Cooley-Tukey method~\cite{cooley-tukey}, though a similar algorithm was also developed much earlier by Gauss~\cite{fbf62979-8f62-38b9-8163-f7084de8b15a}.

A generalization of the Cooley-Tukey method to non-abelian groups which constructs the FFT through subgroup-decomposition was first identified by Diaconis and Rockmore~\cite{diaconis-rockmore}. A naive approach, that would compute Eq.~\eqref{eqn:Fourier-group} directly would yield an asymptotic scaling of $O(\vert G \vert^{2})$. On the other hand, the Diaconis-Rockmore algorithm, like the Cooley-Tukey algorithm, relies on the fact that one could write any irrep of a group element as $\rho(g) = \rho(\ell) \rho(h)$ whenever $g =\ell h$ where $h \in H$ 
belongs to a subgroup of $H \subset G$, and $\ell$ is an element of the left transversal. With this decomposition, the problem essentially reduces to computing the Fourier transform of the subgroup $H$. In this way, Diaconis and Rockmore showed that one can reduce the runtime of computing the Fourier transform of the symmetric group $S_N$ from $O(\vert S_N \vert^{2})$ to $O(\vert S_N \vert \log{\vert S_N \vert})$\footnote{where $|S_N|=N!$}.

Given the unitarity of the Fourier transform, it is possible to implement the DFT and FFT as quantum algorithms. The simplest example is that of simultaneously applying a Hadamard gate to each qubit corresponds to the QFT over the group $\mathbb{Z}^{\otimes n}_2$. The usual quantum Fourier transform (QFT) circuit over the abelian group $\mathbb{Z}_N$ essentially uses this idea with the chain of subgroups $\mathbb{Z}/2^n\mathbb{Z} \supset \mathbb{Z}/2^{n-1}\mathbb{Z} \supset \dots \supset \mathbb{Z}/2\mathbb{Z} \supset \{ \mathbbm{1}\}$~\cite{kitaev,NC,hales2000improved}. Similar QFT circuits can be constructed over other non-abelian groups, as in the early work by Hoyer~\cite{Hoyer:1997qc}, Beals~\cite{beals1997quantum}, and Pueschel et al.~\cite{Pueschel:1998zzo}. The term quantum Fourier transform (QFT) is sometimes used to refer only to the quantum implementation of the FFT for the group $\mathbb{Z}_{2^n}$ but in this work we will use QFT to refer to \emph{any} algorithm for performing Fourier transformations over general finite groups, $G$. For the quantum circuits, we will use $\mathfrak{U}^{G}_{FT}$ and $\mathfrak{U}^{G}_{FFT}$ to represent the naive and fast Fourier transforms respectively. In this case we consider ``naive" to mean explicitly building the unitary matrix corresponding to the Fourier transform and then compiling it using available software libraries, such as Qiskit \cite{QiskitCommunity2017}.

In this work we will construct $\mathfrak{U}^{G}_{FFT}$ for specific discrete subgroups of $SU(2)$ and $SU(3)$ relying on the methods of~\cite{Pueschel:1998zzo}. We briefly review the general construction of fast fourier transforms in Sec.~\ref{sec:subgroupadaptedfouriertransform}.  Sec.~\ref{sec:prim} discusses the basic gates for qubits and qutrits we will use to construct $\mathfrak{U}^{G}_{FFT}$.  The next two sections cover specific $\mathfrak{U}^{G}_{FFT}$s. Sec.~\ref{sec:su2fouriertransformations} is devoted to the crystallike subgroups of $SU(2)$, binary tetrahedral\footnote{It is worth noting that $\mathbb{BT}$ is isomorphic to the Clifford group} ($\mathbb{BT}$) and binary octahedral ($\mathbb{BO}$), and Sec.~\ref{sec:su2fouriertransformations} for the SU(3) subgroups $\Delta(27)$, $\Delta(54)$, and $\Sigma(36\times3)$. Comparison of using $\mathfrak{U}^{G}_{FFT}$ vs $\mathfrak{U}^{G}_{FT}$ in fiducial quantum simulations are presented in Sec.~\ref{sec:outlook}, following by concluding remarks and discussion of future research directions 
in Sec.~\ref{sec:dis}. 

\section{Fast Fourier Transformation Algorithm}
\label{sec:subgroupadaptedfouriertransform}

Before beginning the discussion of the fast Fourier transformation it is crucial to define several technical terms that will be used throughout this work: left (right) transversal, inner conjugate, and left (right) regular representations.
Let $G$ be a group, and $H$ a subgroup. For any $g \in G$, the left (right) cosets of $H$ in $G$ are the sets obtained by left (right) multiplication by $g$. The left (right) transversal for a subgroup $H$ of some group $G$ is then the set of representative elements of each left (right) coset such that exactly one element of each coset appears in the transversal.
An example of this would be numbers $\lbrace 1, i\rbrace$ form the right transversal of the group $\mathbb{Z}_2$ (with elements $\lbrace 1, -1\rbrace$) in $\mathbb{Z}_4$.

A subgroup $H \subset G$ is said to be normal when conjugating any element of $H$ by an element of $G$ returns another element of $H$. In particular, this is true of any element $t$ in the transversal,
\begin{equation}
    t^{-1} h t = h',~\text{with}~ h,h'\in H
\end{equation}
The property of possessing a normal subgroup is a crucial feature of the method of Pueschel et al. \cite{Pueschel:1998zzo}, which allows us to build the Fourier transform for the parent group using this method. An important operation in this regard is the inner conjugation by an element of the transversal $t$ for a given representation, $\phi$, of a subgroup $H$,
\begin{equation}
    \phi^t(h) = \phi(t^{-1} h t).
\end{equation}
Representations for which $\phi^{t}(h) = \phi(h)$ for all $t$ are said to be extendable. When building the Fourier transform for a group $G$ from a normal subgroup $H$, one must build up the irreps of the group $G$ from those of its subgroup $H$, and in the process, the extendable representations incur trivial phase factors.

The final operation we need to consider is the left regular representation. This left regular representation is a functional operator $L(g')$, which takes as input a group element, $g$, and performs the permutation:
\begin{align}
    L(g')|g\rangle = |g'g\rangle,
\end{align}
where $g,g'\in G$. The right regular representation can be similarly defined. The regular representations are important operations as they \emph{must} be block diagonalized by the QFT. In fact, Pueschel et al. \cite{Pueschel:1998zzo} define a QFT as a unitary change of basis of the Hilbert space that block diagonalizes the regular representations.

\begin{figure}[!t]
\centering
\includegraphics[width=\linewidth]{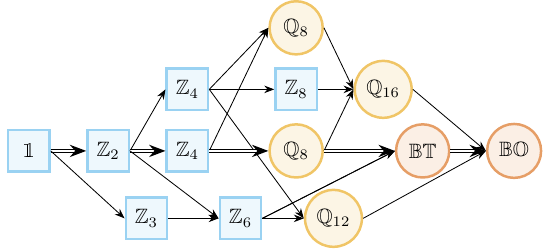}
\caption{Lattice of subgroups of $\mathbb{BO}$.  Each path represents a possible decomposition of $\mathfrak{U}_{FFT}^{G}$.  The double-line path indicates the one explored in this work.}
\label{fig:btpaths}
\end{figure}

Deriving $\mathfrak{U}^{G}_{FFT}$ requires an algorithmic construction through a systematic extension of Fourier transforms along a series of nested subgroups~\cite{Pueschel:1998zzo,Hoyer:1997qc, moore2003generic}.There are numerous paths, given by the lattice of subgroups, one could take to develop FFT for non-abelian groups as shown in Fig.~\ref{fig:btpaths} for the example of $\mathbb{BO}$. Regardless of the path through group space chosen, the general procedure is to find and construct a quantum circuit for a unitary operator which naturally extends the FFT from one finite group in the series to the next \cite{Pueschel:1998zzo, Hoyer:1997qc, moore2003generic}.

\begin{figure}[ht]
\includegraphics[width=\linewidth]{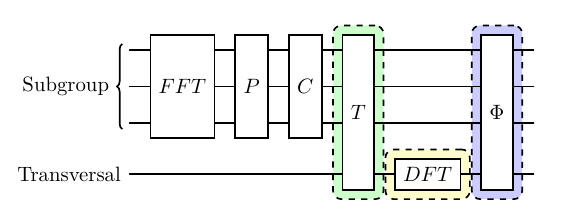}
\caption{Graphical depiction of the $\mathfrak{U}_{FFT}^{G}$ algorithm provided in Ref. \cite{Pueschel:1998zzo}.}
\label{fig:genericfouriertransform}
\end{figure}

While multiple methods for constructing these Fourier transforms exist, we will use the method 
proposed in Ref. \cite{Pueschel:1998zzo}.
The algorithm works by appending a list of elements, denoted a \textit{right transversal}, to an existing group whose QFT is known, e.g. $\mathbb{Z}_4$, and then using the effects of the right transversal to extend the irreps of the subgroup to those of a parent group, e.g. $\mathbb{Q}_8$ or $\mathbb{Z}_8$.  The structure for these circuits is shown in Fig.~\ref{fig:genericfouriertransform}.

In this method, the quantum circuit for a given Fourier transformation can be broken up into six pieces that form a recursive algorithm: the subgroup fast Fourier transform ($FFT$), two irrep permutations ($P$ and $C$), a twiddle matrix ($T$), a cyclic fast Fourier transform over the transversal ($DFT$), and a phase-kickback operation ($\Phi$).

The $FFT$ operation is performed first on the quantum register corresponding to the subgroup.
This operation takes the group element basis into a mixed basis; the subgroup operations exist in the irrep basis while the transversal elements are still in the group element basis. This is the portion of the algorithm that allows for a recursive generation as each successive subgroup can be broken down in this way.

The $P$ and $C$ operations are operations that permute the subgroup irreps such that the book-keeping and quantum circuits are easily manipulated. $P$ takes irreps and permutes them in such a way that they are identifiable for easy bookkeeping.
The $C$ operation then takes irreps which are not invariant under conjugation by the transversal, i.e. $\rho^{a}(t g t^{-1}) \neq \rho^{a}$, and adjusts them so they are nested together.
This will be important later for the phase-kickback operation, $\Phi$. 

The twiddle operation, $T$ is an operation controlled on the transversal register that either mixes the conjugate sets to make a larger dimensional irrep or transforms an existing $n$-dimensional irrep under the explicit representation. This is immediately followed by $DFT$ on the transversal register which mixes the irreps and ensures that the transversal component is moved to the irrep basis. The final step involves a phase kickback that applies certain phases to the given irreps so that they are the actual irreps and not an isomorphism. This is important as even though irreps are really only meaningful within the Fourier transform up to unitary equivalence, in the block diagonalized regular representation, the same irrep must appear as many times as its dimensionality along the block-diagonal. These pieces together form the recursive structure necessary to build a nonabelian quantum Fourier transform.

In this work we construct $\mathfrak{U}_{FFT}^G$ for five cases not currently provided in the literature.  These are $\mathbb{BT},\mathbb{BO}$ and $\Delta(27),\Delta(54),\Sigma(36\times3)$ which correspond to subgroups of $SU(3)$ and $SU(2)$ respectively.  The nested subgroup paths we will construct are:
\begin{align}
    & \mathbbm{1}\rightarrow \mathbbm{Z}_3\times\mathbbm{Z}_3 \rightarrow \Delta(27)\rightarrow\Delta(54)\rightarrow\Sigma(36\times3)\\
            & \mathbbm{1} \rightarrow \mathbbm{Z}_2 \rightarrow \mathbbm{Z}_4 \rightarrow \mathbbm{Q}_8 \rightarrow \mathbbm{BT} \rightarrow \mathbbm{BO}.\label{eq:btseries}
\end{align}
The $\mathfrak{U}_{FFT}^G$ for $\mathbb{BT}$, $\mathbb{BO}$, and $\Sigma(36\times3)$ provide the resource cost reductions for quantum simulations of lattice gauge theories for these respective groups~\cite{Gustafson:2022xdt, Gustafson:2023kvd, gustafson2024primitive}.

\section{Primitive Gates}
\label{sec:prim}
Throughout this work we consider two quantum architectures: the standard qubit-based and a heterogeneous qubit-qutrit one.  Our consideration of qubit-qutrit devices is motivated by the non-abelian groups studied here, which contain parts that are naturally trinary-valued. In this section, we discuss the primitive gates we will use to construct the FFTs.  

The qubit gates include the usual arbitrary $U(2)$ rotations (including $X,Y,Z$), $CNOT$, and the 3-qubit Toffoli. While $CNOT$+U(2) forms a universal set \cite{Chuang:1996hw}, we use Toffoli for compression of circuits in presenting them.  Ultimately, all these circuits will be decomposed into a fault-tolerant gate set. For most cases, we will decompose all the operations to the lowest possible form; the primary exception will be arbitrary two-qubit rotations for which efficient decompositions are readily available~\cite{barenco1995elementary,NC}. 

For qubit-qutrit gates, we use the set of gates provided in Ref. \cite{gustafson2024primitive}. The one qutrit gates include the generalization of the Pauli gates to qutrits,
\begin{align}
    X_{a,b} &= |a\rangle \langle b| + |b\rangle \langle a| + \sum_{c\neq a,b} |c\rangle \langle c|\notag\\
    Z_{a} &= \sum_{c=0}^2 (1 - 2 \delta_{c, a}) |c\rangle \langle c|,
\end{align}
and the qutrit equivalent of the Hadamard gate,
\begin{align}
\label{eq:dah3gate}
    H_3 = \frac{1}{\sqrt{3}} \begin{pmatrix}
        1 & 1 & 1\\
        1 & \omega_3 & \omega_3^2\\
        1 & \omega_3^2 & \omega_3\\
    \end{pmatrix},
\end{align}
where $\omega_3=e^{2\pi i / 3}$. Similar to the qubit Hadamard, this corresponding to a FFT on $\mathbb{Z}_3$. Another important single qutrit operation is implemented by the clock-shift gate and its inverse
\begin{align}
\label{eq:clockshiftmatrix}
    \chi = \begin{pmatrix}
        0 & 1 & 0 \\
        0 & 0 & 1 \\
        1 & 0 & 0 \\
    \end{pmatrix},\qquad \chi^{-1} = \begin{pmatrix}
        0 & 0 & 1 \\
        1 & 0 & 0 \\
        0 & 1 & 0 \\
    \end{pmatrix} .
\end{align}

The final qutrit gates are the $T_3$ and $S_3=T_3^3$ which are the generalizations of the $T$ and $S$ gates for qubits that are important for fault-tolerant compilation:
\begin{equation}
    T_3=\text{Diag}(1,\omega_9,\omega_9^8),\quad S_3=\text{Diag}(1,\omega_3,\omega_3^2)
\end{equation}
where $\omega_n=e^{2\pi i/n}$ is the $n-$th root of unity.

For entangling gates, we consider the controlled sum (controlled increment) gate which performs modular addition on the qutrits and is denoted as a controlled $\chi$. For each of these qutrit operators, we also desire a qubit decomposition. 

Because of the mismatch in dimensionality between qutrits and qubits, there is some freedom in how to embed a single qutrit in two qubits. We choose to assign states according to their decimal integer values, such that 
\begin{align}
    |0\rangle_3 \equiv |00\rangle_2,\qquad |1\rangle_3 \equiv |01\rangle_2 ,\qquad |2\rangle_3 \equiv |10\rangle_2.
\end{align}
and the $|11\rangle_2$ state is forbidden.  This embedding is nice because it allows for easy construction of the two-qubit unitary from the qutrit one by simply direct summing the qutrit unitary with $\mathbbm{1}_1$.  For example, the qubit $H_3$ is

\begin{equation}
    \label{eq:H3examp}
    H^{\rm qubit}_3 = H_3^{\rm qutrit}\oplus  \mathbbm{1}_1=\frac{1}{\sqrt{3}}\begin{pmatrix} 1 & 1 & 1 & 0\\
    1 & \omega_3 & \omega_3^2 & 0\\
    1 & \omega_3^2 & \omega_3 &0 \\
    0 & 0 & 0 & \sqrt{3}
    \end{pmatrix}.
\end{equation}
This unitary can be decomposed using available transpilers such as \textsc{Qiskit} to 3 CNOTs and 14 $R_z$ gates. We also denote an additional $H_3$ gate mapping for qubits which appears in the derivation of the $\mathbb{BT}$ circuits; this operator is
\begin{equation}
   (H_{3}^{\rm qubit})' = \mathbbm{1}\oplus H_{3}^{\rm qutrit} = \frac{1}{\sqrt{3}}\begin{pmatrix} \sqrt{3} & 0 & 0 & 0 \\
    0 & 1 & 1 & 1\\
    0 & 1 & \omega_3 & \omega_3^2\\
    0 & 1 & \omega_3^2 & \omega_3
    \end{pmatrix}.
\end{equation}

The qubit version of the permutation gate $\chi$ defined in Eq. (\ref{eq:clockshiftmatrix}) is provided in Fig. \ref{fig:clockmatrixqubit}. If this operator needs to be controlled then one applies this operator once provided the product of the controls is $1$ modulus 3, the inverse operator is applied if the product of the controls is $2$ modulus 3, and left alone if the product of controls is $0$ modulus 3. The most efficient way to do this for more than one control is to compute the modulo 1 and modulo 2 values into two ancillae and then apply $\chi$ or $\chi^\dagger$ controlled on the respective ancilla. 

\begin{figure}[t]
\begin{quantikz}[row sep=1em]
    & \gate[1]{\chi} & \qw 
\end{quantikz}
 $\equiv$
\begin{quantikz}
    & \targ{} & \ctrl{1} & \targ{} & \qw \\
    & \ctrl{-1} & \targ{} & \qw & \qw 
\end{quantikz}
\caption{Decomposition of the qutrit $\chi$ gate onto a 2 qubit register.
}
\label{fig:clockmatrixqubit}
\begin{quantikz}[row sep=0em]
& \gate[1]{Z_2} & \qw \\
\end{quantikz} = 
\begin{quantikz}[row sep=2em]
    & \ctrl[open]{1} & \qw \\
    & \control{} & \qw 
\end{quantikz}~~
\begin{quantikz}[row sep=0.em]
    & \gate[1]{X_{1,2}} & \qw \\
\end{quantikz} =
\begin{quantikz}[row sep=2.2em]
    & \targX{} & \qw \\
    & \targX{} \wire[u]{q} & \qw 
\end{quantikz}
\caption{Circuit Decomposition of qutrit $Z_2$ gate onto a two qubit register (left) and decomposition of the qutrit $X_{1,2}$ gate onto a two qubit register (right).
}
\label{fig:z2qubit}
\end{figure}

The two remaining circuit operators that need decomposition are the phase operation $Z_2$ and the permuation operator $X_{1,2}$. The phase operation is given by
\begin{align}
    Z_2 = \text{Diag}(1,1,-1,1),
\end{align} with its circuit found in Fig.~\ref{fig:z2qubit}. The $X_{1,2}$ gate is
\begin{align}
    X_{1,2} = \begin{pmatrix}
        1 & 0 & 0 & 0\\
        0 & 0 & 1 & 0\\
        0 & 1 & 0 & 0\\
        0 & 0 & 0 & 1\\
    \end{pmatrix},
\end{align}
and is identifiable as the usual $SWAP$ gate (See Fig. \ref{fig:z2qubit}).

A final crucial component for transforming qutrit gates to qubit ones is how to implement control on the resulting multi-qubit states. In theory, this just involves on controlling on the states $|00\rangle_2$, $|01\rangle_2$, and $|10\rangle_2$ respectively. However, because the $|11\rangle_2$ state is forbidden, some control operations can be simplified. Specifically, for the $|01\rangle_2$ and $|10\rangle_2$ states, one only needs to control on the second or first register respectively because the $|11\rangle$ state is \emph{never} occupied. When necessary, we also use the qudit $|d\rangle$ notation when considering pure matrix representations and the qubit $|ab\rangle$ notation when mapping to quantum circuits. Now that we have cleared this ambiguity.

These rules are sufficient for decomposing the circuits provided in the main text.

\section{QFT for Subgroups of \texorpdfstring{$SU(2)$}{SU(2)}}
\label{sec:su2fouriertransformations}
As discussed above, there are numerous paths to construct an FFT for $\mathbb{BT}$ and $\mathbb{BO}$.  We will pursue the path 
\begin{align}
\label{eq:btpath}
1 \rightarrow \mathbb{Z}_2\rightarrow \mathbb{Z}_4 \rightarrow \mathbb{Q}_8 \rightarrow \mathbb{BT} \rightarrow \mathbb{BO}
\end{align}
as it correspond most closely to the results from \cite{Gustafson:2022xdt, Gustafson:2023kvd}. 
The consequence of this choice will become apparent following the discussion of the group structures, their presentations in terms of generating elements, and irreps.
\subsection{Structure and irreps of \texorpdfstring{$\mathbb{Q}_8$, $\mathbb{BT}$, and $\mathbb{BO}$}{}}

Before delving into the Fourier transformations, it is important to list out the structure of the principal finite groups of interest for SU(2): $\mathbb{Q}_8$, $\mathbb{BT}$, and $\mathbb{BO}$. 
Each of these groups forms a subgroup in a nested series which extend from one subgroup to the next via a right transversal.
The groups $\mathbb{Q}_8$, $\mathbb{BT}$, and $\mathbb{BO}$ can respectively be represented with the following ordered products
\begin{align}
f = &(-1)^a \mathbf{j}^b \mathbf{k}^c \\
g = &(-1)^a \mathbf{j}^b \mathbf{k}^c \mathbf{u}^d\\
 h = &(-1)^a\mathbf{j}^b\mathbf{k}^c\mathbf{u}^d\mathbf{t}^e.
\end{align}
The generators which appear in these above equations have the following matrix presentation in the faithful representation:
\begin{align}
\label{eq:btops}
    \mathbf{j} &= \begin{pmatrix}
    0 & 1\\
    -1 & 0\\
    \end{pmatrix},~
    \mathbf{k} = \begin{pmatrix}
        i & 0\\
        0& -i\\
    \end{pmatrix},\notag\\
    \mathbf{u}&=\begin{pmatrix} -\eta & -\eta\\
   \eta^* & -\eta^*
    \end{pmatrix},~
     \mathbf{t} = \frac{1}{\sqrt{2}}\begin{pmatrix} 1 & -i\\
                                                     -i & 1\\
                                     \end{pmatrix}
\end{align}
where $\eta = \frac{1 + i}{2}$, $0\leq a, b, c, e\leq 1$, and $0\leq d \leq 2$.     

Additional details about $\mathbb{BT}$ and $\mathbb{BO}$ can be found in Refs. \cite{Gustafson:2022xdt, Gustafson:2023kvd} however we quote here the important information regarding the irreps, denoted $\xi^m$ for $\mathbb{Q}_8$, $\rho^m$ for $\mathbb{BT}$ and $\bar{\rho}^m$ for $\mathbb{BO}$. The irreps for all three groups are listed in Tabs. \ref{tab:q8irreps}, \ref{tab:btirreps}, and \ref{tab:boirreps} with $\omega_3 = e^{2\pi i / 3}$. 

\begin{table}
\caption{Matrix representations of the $\mathbb{Q}_8$ generators. The rows indicate the irrep while the columns indicate the generator. Further $\mathbf{j}_2,\mathbf{k}_2$ are the 2d matrix representations in Eq.~(\ref{eq:btops}).}
\label{tab:q8irreps}
\centering
\begin{tabular}{c|ccc}
$f$ & $-1$ & $\mathbf{j}$ & $\mathbf{k}$\\\hline\hline
$\xi^1$ & 1 & 1 & 1\\
$\xi^2$ & 1 & $-1$ & 1\\
$\xi^3$ & 1 & 1 & $-1$\\
$\xi^4$ & 1 & $-1$ & $-1$\\
$\xi^5$ & $-\mathbbm{1}_2$ & $\mathbf{j}_2$ & $\mathbf{k}_2$
\end{tabular}
\end{table}

\begin{table}
    \caption{Matrix representations of the $\mathbb{BT}$ generators. The rows indicate the irrep while the columns indicate the generators. For compactness, $\omega_3 = e^{2\pi i / 3}$ and $\mathbf{j}_2,\mathbf{k}_2,\mathbf{u}_2$ are the 2d matrix representation in Eq.~(\ref{eq:btops}).}
    \label{tab:btirreps}
    \centering
    \begin{tabular}{c|ccccc}
         g&  $-1$&  $\mathbf{j}$&  $\mathbf{k}$&  $\mathbf{u}$\\
         \hline\hline
         $\rho^1$ &  1&  1&  1&  1\\
         $\rho^2$&  1&  1&  1&  $\omega_3$\\
         $\rho^3$&  1&  1&  1&  $\omega_3^2$\\
         $\rho^4$&  $-\mathbbm{1}_2$ & $\mathbf{j}_2$ & $\mathbf{k}_2$ & $\mathbf{u}_2$\\ 
         $\rho^5$&  $-\mathbbm{1}_2$ & $\mathbf{j}_2$ & $\mathbf{k}_2$  & $\omega_3\mathbf{u}_2$\\ 
         $\rho^6$&  $-\mathbbm{1}_2$ & $\mathbf{j}_2$ & $\mathbf{k}_2$  & $\omega_3^2\mathbf{u}_2$\\
         $\rho^7$& $\mathbbm{1}_3$ & $\text{Diag}(-1,1,-1)$& $\text{Diag}(1,-1,-1)$
    &$\chi$
    \end{tabular}
    \end{table}

    \begin{table*}
    \caption{Matrix representations of the generators used for digitization of $\mathbb{BO}$. The rows of the table correspond to the irreps while the columns correspond to the group element in the given representation. For compactness, $\omega_3 = e^{2\pi i / 3}$ and $\mathbf{j}_2,\mathbf{k}_2,\mathbf{u}_2,\mathbf{t}_2$ are the 2d matrix representations in Eq.~(\ref{eq:btops}).}
    \label{tab:boirreps}
    \centering
    \begin{tabular}{c|ccccc}
         h&  -1&  $\mathbf{j}$&  $\mathbf{k}$&  $\mathbf{u}$& $\mathbf{t}$\\
         \hline\hline
         $\bar{\rho}_1$ &  1&  1&  1&  1& 1\\
         $\bar{\rho}_2$&  1&  1&  1&  1& -1\\
         $\bar{\rho}_3$&  $\mathbbm{1}_2$ & $\mathbbm{1}_2$ & $\mathbbm{1}_2$ & $\text{Diag}(\omega_3^2,\omega_3)$ & $X$\\
         $\bar{\rho}_4$& $-\mathbbm{1}_2$  & $\mathbf{j}_2$ & $\mathbf{k}_2$  & $\mathbf{u}_2$ 
    & $\mathbf{t}_2$ \\
         $\bar{\rho}_5$& $-\mathbbm{1}_2$  & $\mathbf{j}_2$ & $\mathbf{k}_2$  & $\mathbf{u}_2$ 
    & $-\mathbf{t}_2$\\
    
         $\bar{\rho}_6$&  $\mathbbm{1}_3$  & $\text{Diag}(-1,1,-1)$ & $\text{Diag}(1,-1,-1)$ & $\chi$ & $-(\mathbf{u}_2\otimes \mathbbm{1}_1)$\\
         
    $\bar{\rho}_7$& $\mathbbm{1}_3$ & $\text{Diag}(-1,1,-1)$ &$\text{Diag}(1,-1,-1)$ & $\chi$ & $(\mathbf{u}_2\otimes \mathbbm{1}_1)$\\     
         $\bar{\rho}_8$&  $-\mathbbm{1}_4$ & $\begin{pmatrix}
             0 & -i & 0 & 0 \\
             -i & 0 & 0 & 0\\
             0 & 0 & -i & 0\\
             0 & 0 & 0 & i\\
         \end{pmatrix}$ & $\begin{pmatrix}
             i & 0 & 0 & 0 \\
             0 & -i & 0 & 0\\
             0 & 0 & 0 & -i\\
             0 & 0 & -i & 0\\
         \end{pmatrix}$ & $\omega_3\begin{pmatrix}
    -\eta\omega_3 & \eta\omega_3 & 0 & 0 \\
    -\eta^*\omega_3 & -\eta^*\omega_3 & 0 & 0 \\
    0 & 0 & -\eta^* & \eta \\
    0 & 0 & -\eta^*& -\eta
\end{pmatrix}$ & $\begin{pmatrix}
    0 & 0 & 0 & -1\\ 0 & 0 & 1 & 0 \\ 1 & 0 & 0 & 0 \\ 0 & 1 & 0 & 0
\end{pmatrix}$\\
    \end{tabular}
\end{table*}

\subsection{Fourier Transformation for \texorpdfstring{\BT}{}}
\label{sec:btfouriertransform}
\begin{figure}[b]

\centering
\includegraphics[width=0.7\linewidth]{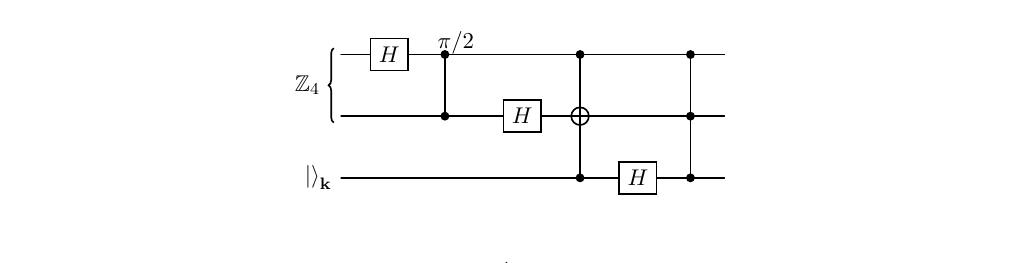}
\caption{Quantum circuit for $\mathfrak{U}_{FFT}^{\mathbb{Q}_8}$ from Ref.~\cite{Pueschel:1998zzo}.
}
\label{fig:Q8fft}
\end{figure} 

The construction of $\mathfrak{U}_{FFT}^{\mathbb{BT}}$ requires the identification of the six constituent gates: $\mathbb{Q}_8$ $FFT$, $P$, $C$, $T$, $DFT$, and $\Phi$.
The first, the $FFT_{\mathbb{Q}_8}$ was derived in~\cite{Pueschel:1998zzo} and is presented in  Fig. \ref{fig:Q8fft}. Given the choice of implementation for $\mathcal{T}_{\mathbb{BT}}$ that will be taken below, we will see that the $C$ and $P$ transformations are unnecessary.

We thus need to connect the irrep vectors of $\mathbb{Q}_8$, $|\xi^a\rangle$, with their corresponding computational basis states.
These states are explicitly
\begin{align}
|0\rangle_{\mathbf{k}}|0\rangle_{\mathbf{j}}|0\rangle_{-1} =  |\xi^1\rangle, ~~ |1\rangle_{\mathbf{k}}|0\rangle_{\mathbf{j}}|0\rangle_{-1} = |\xi^3\rangle\notag\\ 
|0\rangle_{\mathbf{k}}|1\rangle_{\mathbf{j}}|0\rangle_{-1} =  |\xi^2\rangle,~~|1\rangle_{\mathbf{k}}|1\rangle_{\mathbf{j}}|0\rangle_{-1} = |\xi^4\rangle\notag\\
\lbrace |a\rangle_{\mathbf{k}}|b\rangle_{\mathbf{j}}|1\rangle_{-1}: 0\leq a,b\leq 1\rbrace =  |\xi^5\rangle,
\end{align}
where the corresponding irreps are enumerated in Table \ref{tab:q8irreps}. The only remaining step to generate the other gates is to identify the sets of inner conjugates. For $\mathbb{BT}$, we can identify the inner conjugacy transformations using the results from Refs. \cite{Gustafson:2022xdt, Gustafson:2023kvd}:
\begin{align}
\mathbf{u} (-1)^a \mathbf{j}^b \mathbf{k}^c \mathbf{u}^2 = & (-1)^a \mathbf{j}^{c}\mathbf{k}^{b + c} \notag\\
 \mathbf{u}^2 (-1)^a \mathbf{j}^b \mathbf{k}^c \mathbf{u} = & (-1)^a \mathbf{j}^{b + c} \mathbf{k}^b.
\end{align}
These transformation rules immediately tell us what the pairs of inner conjugates are; $\xi^1$ and $\xi^5$ are left invariant under this transformation, i.e. these irreps are extendable. 
Meanwhile, we can identify that $(\xi^2)^{\mathbf{u}}(f) = \xi^4(f)$ and $(\xi^2)^{\mathbf{u}^2}(f) = \xi^3(f)$. This implies that $\lbrace \xi^2,~\xi^4,~\xi^3\rbrace$ form a set of inner conjugates and are not extendable representations.

\begin{figure}
    \includegraphics[width=\linewidth]{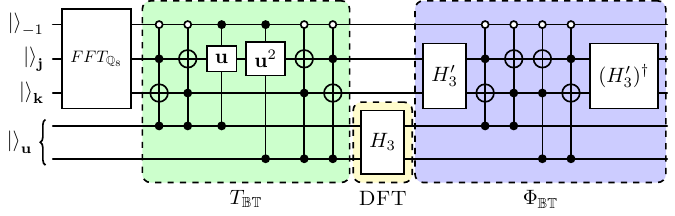}
    \caption{Quantum circuit for $\mathfrak{U}_{FFT}^{\mathbb{BT}}$. The gate $H_3'$ is a discrete Fourier transformaiton over $\mathbb{Z}_3$ on the $|01\rangle$, $|10\rangle$, $|11\rangle$ register. }
    \label{fig:BTFFT}
\end{figure}

The irreps $\lbrace \xi^2,~\xi^3,~\xi^4\rbrace$ when combined with the transversal, $\lbrace 1,~\mathbf{u},~\mathbf{u}^2\rbrace$, make the 3d irrep of $\mathbb{BT}$, $\rho^7$. 
This allows us to write down the twiddle operation $T$. This operation will end up being a controlled operation on the $\mathbb{Q}_8$ register of the form 
\begin{equation}
\label{eq:twidBT}
T_{\mathbb{BT}} = \sum_{d=0}^2 |d\rangle\langle d| \otimes \mathcal{T}_{\mathbb{BT}}^d
\end{equation}
where
\begin{align}
\mathcal{T}_{\mathbb{BT}} = \mathbbm{1}_{1} \oplus \chi \oplus (\mathbf{u}_2\otimes \mathbbm{1}_2).
\end{align}
The quantum circuit which implements this transformation is highlighted in green in Fig. \ref{fig:BTFFT}. The first two Toffoli gates perform the cyclic permutations, while the controlled $\mathbf{u}$ operation performs a controlled rotation on the $|\rangle_{\mathbf{j}}$ register corresponding to the presentation of $\mathbf{u}$ in the $\rho^4$ irrep show in Eq.~\ref{eq:btops}.
 This leaves the implementation of $\Phi$ which can be broken down similar to $T_{\mathbb{BT}}$ in Eq. (\ref{eq:twidBT}) as
\begin{align}
\Phi_{\mathbb{BT}} = \sum_{d=0}^2 |d\rangle\langle d| \otimes \phi_{\mathbb{BT}}^d,
\end{align}
where
\begin{align}
\phi_{\mathbb{BT}} = \text{Diag}(1,1, \omega, \omega^2, 1, 1, 1, 1).
\end{align}
The phase kickback here needs to be applied to the irreps which form a conjugate set, in this case $\xi^2$, $\xi^3$, $\xi^4$ but leaves the other irreps unchanged. 
The complete quantum Fourier transformation for \BT~then strings these gates together and is shown in Fig.~\ref{fig:BTFFT} with the phase kickback highlighted in blue. It is important to note the following, the $H_3'$ gates do not need to be controlled on the $\ket{}_{\mathbf{u}}$ register since it forms a basis transformation taking the permutations of the Toffoli gates to phases. 

\subsection{Fourier Transformation for \texorpdfstring{\BO}{}}

The construction of the Fourier transformation for \BO~will follow that of \BT, with the subgroup FFT being that of \BT.  Next, we can identify the irreps after $\mathfrak{U}_{FFT}^{\mathbb{BT}}$ with their corresponding basis vectors. 
The 1d irreps correspond to the basis vectors
\begin{align}
|0\rangle_{\mathbf{u}}|0\rangle_{\mathbf{k}}|0\rangle_{\mathbf{j}}|0\rangle_{-1} &= |\rho^1\rangle \notag\\
 |1\rangle_{\mathbf{u}}|0\rangle_{\mathbf{k}}|0\rangle_{\mathbf{j}}|0\rangle_{-1} &= |\rho^2\rangle\notag\\
 |2\rangle_{\mathbf{u}}|0\rangle_{\mathbf{k}}|0\rangle_{\mathbf{j}}|0\rangle_{-1} &= |\rho^3\rangle.
\end{align}
The 2d irreps corresponding to sets of quadruplets
\begin{align}
    \lbrace |0\rangle_{\mathbf{u}} |a\rangle_{\mathbf{k}}|b\rangle_{\mathbf{j}}|1\rangle_{-1} \rbrace &= |\rho^4\rangle\notag\\
    \lbrace |1\rangle_{\mathbf{u}} |a\rangle_{\mathbf{k}}|b\rangle_{\mathbf{j}}|1\rangle_{-1} \rbrace &= |\rho^5\rangle\notag\\
    \lbrace |2\rangle_{\mathbf{u}} |a\rangle_{\mathbf{k}}|b\rangle_{\mathbf{j}}|1\rangle_{-1} \rbrace &= |\rho^6\rangle
\end{align} 
where $0\leq a,b\leq 1$. 
Finally the 3d irrep will correspond to the nonaplet
\begin{align}
\lbrace |d\rangle_{\mathbf{u}} |1\rangle_{\mathbf{k}}|0\rangle_{\mathbf{j}}|0\rangle_{-1}, &\notag\\
       |d\rangle_{\mathbf{u}} |0\rangle_{\mathbf{k}}|1\rangle_{\mathbf{j}}|0\rangle_{-1}, &\notag\\
       |d\rangle_{\mathbf{u}} |1\rangle_{\mathbf{k}}|1\rangle_{\mathbf{j}}|0\rangle_{-1}\rbrace & = |\rho^7\rangle
\end{align}
for $0\leq d \leq 2$. We next identify the inner conjugates under $\mathbf{t}$ from the expression
\begin{equation}
\mathbf{t} (-1)^a \mathbf{j}^b \mathbf{k}^c \mathbf{u}^d \mathbf{t}^{-1} = 
(-1)^{a'} \mathbf{j}^{c + \delta_{d,1} + \delta_{d,2}} \mathbf{k}^{b + \delta_{d, 1}} \mathbf{u}^{2d}
\end{equation}
where $a'$ is an unimportant polynomial of the $c$, $b$, $\delta_{d,1}$, and $\delta_{d,2}$. 
We note that the nontrivial effect of inner conjugation by $\mathbf{t}$ is $\mathbf{u} \leftrightarrow \mathbf{u}^2$.
It then follows that  
\begin{align}
(\rho^1)^\mathbf{t} = \rho^1,\quad (\rho^4)^{\mathbf{t}}&= \rho^4,\quad (\rho^7)^{\mathbf{t}} = \rho^7\notag\\
(\rho^2)^{\mathbf{t}} &= \rho^3\notag\\
(\rho^5)^{\mathbf{t}} &= \rho^6
\end{align}
and thus the inner conjugates are $\{\rho^2, \rho^3\}$ and $\{\rho^5,\rho^6\}$. This implies that $\rho^2$ and $\rho^3$ will mix to create the 2d irrep of $\mathbb{BO}$, $\bar{\rho}_3$. Meanwhile the 2d irreps $\rho^5$ and $\rho^6$ will combine to generate the 4d irrep $\bar{\rho}_8$.

\begin{figure}
    \includegraphics[width=\linewidth]{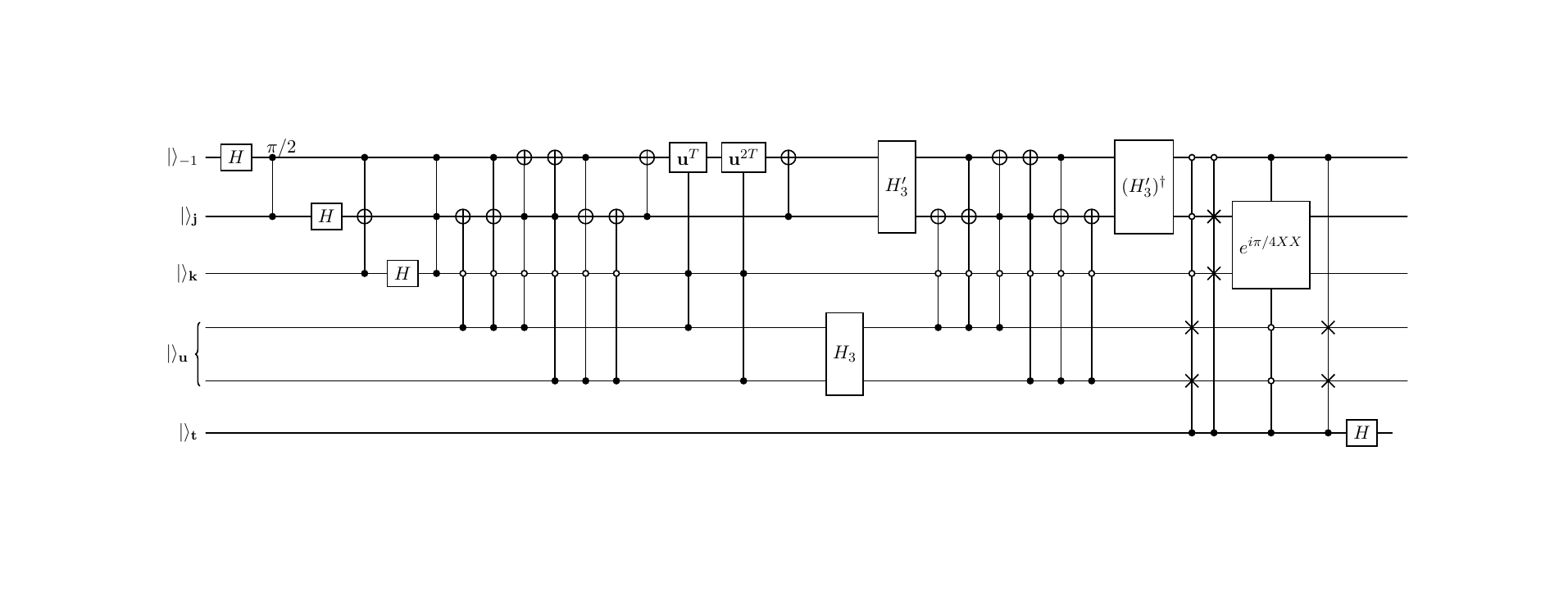}
    \caption{Quantum circuit for $\mathfrak{U}_{FFT}^{\mathbb{BO}}$. The twiddle and phase kickback operations are highlighted in green and blue respectively.}
    \label{fig:BOFFT}
\end{figure}

Everything is now in place to construct the twiddle operator, $T_{\mathbb{BO}}$, and the phase kickback $\Phi_{\mathbb{BO}}$.
Similar to \BT, we can write decompose the twiddle matrix
\begin{align}
    T_{\mathbb{BO}} = \sum_{e=0}^1 |e\rangle\langle e| \otimes \mathcal{T}_{\mathbb{BO}}^e.
\end{align}
The only remaining issue is to identify what the matrix $\mathcal{T}_{\mathbb{BO}}$ is. We can identify this by evaluating the effects on the basis states for $\mathbb{BT}$:

\begin{align}
\label{eq:twiddleBO}
\mathcal{T}_{\mathbb{BO}} &= \mathbbm{1}_{1} \oplus X\oplus (\mathbbm{1}_{3}\otimes [-(\mathbf{u}_2\otimes\mathbbm{1}_1)])\oplus\notag\\
&\begin{pmatrix}
\frac{1}{\sqrt{2}} & 0 & 0 & 0 & 0 & 0 & \frac{-i}{\sqrt{2}} &  0 & 0 & 0 & 0 & 0\\
0 & 0 & 0 & 0 & 0 & 0 & 0 & 0 & 0 & 0 & 1 & 0\\
0 & 0 & 0 & 0 & 0 & 0 & 0 & 0 & 0 & 0 & 0 & 1\\
0 & 0 & 0 & \frac{1}{\sqrt{2}} & 0 & 0 & 0 & 0 & 0 & \frac{-i}{\sqrt{2}} &  0 & 0\\
0 & 0 & 0 & 0 & 0 & 0 & 0 & -1 & 0 & 0 & 0 & 0\\
0 & 0 & 0 & 0 & 0 & 0 & 0 & 0 & -1 & 0 & 0 & 0\\
\frac{-i}{\sqrt{2}} & 0 & 0 & 0 & 0 & 0 & \frac{1}{\sqrt{2}} & 0 & 0 & 0 & 0 & 0\\
0 & 1 & 0 & 0 & 0 & 0 & 0 & 0 & 0 & 0 & 0 & 0\\
0 & 0 & 1 & 0 & 0 & 0 & 0 & 0 & 0 & 0 & 0 & 0\\
0 & 0 & 0 & \frac{-i}{\sqrt{2}} & 0 & 0 & 0 & 0 & 0 & \frac{1}{\sqrt{2}} & 0 & 0\\
0 & 0 & 0 & 0 & 1 & 0 & 0 & 0 & 0 & 0 & 0 & 0\\
0 & 0 & 0& 0 & 0 & 1 & 0 & 0 & 0 & 0 & 0 & 0\\
\end{pmatrix}
\end{align} 
This operator can be translated to a quantum circuit with a modicum of work into the one shown in the green highlighted portion of Fig.~\ref{fig:BOFFT}. In most cases the ancilla present in the quantum circuit will be unentangled either through running the Fourier transformation in reverse or by otherwise uncomputing the values they hold.

The only remaining component necessary is the phase-kickback
\begin{align}
    \Phi_{\mathbb{BO}} = \sum_{e=0}^1|e\rangle\langle e|\otimes \phi^d_{\mathbb{BO}}
\end{align}
The phase kickback gets applied to the original states corresponding to $|\rho^3\rangle$ and $|\rho^6\rangle$ irreps.
This yields the operator 
\begin{align}
\label{eq:bophase}
    \phi_{\mathbb{BO}} = & \text{Diag}(1, 1, -1) \oplus \mathbbm{1}_{9}\oplus \notag\\
    & \text{Diag}(1, 1, -1, 1, 1, -1, 1, 1, -1, 1, 1, -1).
\end{align}
We provide the full quantum Fourier transformation for $\mathbb{BO}$ in Fig. \ref{fig:BOFFT}. This circuit includes 2 ancilla which are eventually uncomputed when the transformation is run in reverse. 

\section{QFT for Subgroups of \texorpdfstring{$SU(3)$}{SU(3)}}
\label{sec:su3fouriertransformations}
For the case of $SU(3)$, there are two motivations for the particular groups we investigate. So-called \textit{flavour groups} are important for model-building in particle physics to understand the mass splittings between different flavor generations, e.g. see Ref. \cite{luhn2007flavor}. $\Delta(3n^2)$ and $\Delta(6n^2)$ are prime examples of flavour groups. Other groups, for example $\Sigma(\phi\times3)$, have found interest for simulations in quantum simulations for lattice gauge theorys in HEP~\cite{gustafson2024primitive}. For the work here, it is nice that $\Delta(3n^2)$ form normal subgroups of $SU(3)$

\subsection{Structure and irreps of \texorpdfstring{$\Delta(27)$, $\Delta(54)$, and $\Sigma(36\times3)$}{}}

The group elements, $g \in \Delta(27)$, $h \in \Delta(54)$, and $f \in \Sigma(3\times 36)$, can be written as
\begin{align}
    \label{eq:s108pres1}
    g &= \omega_3^p \mathbf{C}^q \mathbf{E}^r\\
    h &= \omega_3^p \mathbf{C}^q \mathbf{E}^r \mathbf{V}^{2s}\\
    \label{eq:s108pres3}
    f &= \omega_3^p \mathbf{C}^q \mathbf{E}^r \mathbf{V}^{2s + t}
\end{align}
respectively, where $0\leq p,~q,~r \leq 2$ and $0\leq s,~t\leq 1$~\cite{Grimus:2010ak}. 
Each of these subgroups is extended to a larger subgroup of $SU(3)$ according to
\begin{equation}
    \mathbbm{1}\rightarrow \mathbbm{Z}_3\times\mathbbm{Z}_3 \xrightarrow{\mathbf{E}} \Delta(27)\xrightarrow{\mathbf{V}^2}\Delta(54)\xrightarrow{\mathbf{V}}\Sigma(36\times3)
\end{equation}
where the elements above the arrows indicate the right transveral used to extend the group.
We also provide the irreps for $\Delta(27)$, $\Delta(54)$, and $\Sigma(3\times 36)$ are denoted $\xi^a$, $\bar{\rho}^a$, and $\rho^a$ respectively and provided in Tabs. \ref{tab:su3irrepsd27}, \ref{tab:su3irrepsd54}, and \ref{tab:su3irrepss108}.
In addition, we also need to enumerate the irreps of $\mathbb{Z}_3\times\mathbb{Z}_3$. We denote these irreps as $\chi^{3a + b + 1}$ with $0\leq a, b\leq 2$. These irreps have the following presentation
    \begin{align}
        \chi^{3a + b + 1}(\omega_3^p \mathbf{C}^q) = e^{2\pi i / 3 (a p + q b)}
    \end{align}
    where $p$ and $q$ correspond to the generators $\omega_3$ and $\mathbf{C}$ in Eqs.~(\ref{eq:s108pres1}) to (\ref{eq:s108pres3}).
    
\begin{table}
\caption{The 11 irreps for $\Delta(27)$ for each of the generators, $\omega_3$, $\mathbb{C}$, and $\mathbb{E}$.}
\label{tab:su3irrepsd27}
    \begin{tabular}{c|ccc}
    g & $\omega_3$ & $\mathbf{C}$ & $\mathbf{E}$ \\\hline\hline
    $\xi^1 $ & 1 & 1 & 1  \\
    $\xi^2 $ & 1 & $\omega_3 $& 1  \\
    $\xi^3 $ & 1 & $\omega_3^2$ & 1  \\
    $\xi^4 $ & 1 & 1 & $\omega_3 $ \\
    $\xi^5 $ & 1 & $\omega_3 $& $\omega_3 $ \\
    $\xi^6 $ & 1 & $\omega_3^2$ & $\omega_3 $\\
    $\xi^7 $ & 1 & 1 & $\omega_3^2$ \\
    $\xi^8 $ & 1 & $\omega_3 $& $\omega_3^2$ \\
    $\xi^9 $ & 1 & $\omega_3^2$ & $\omega_3^2$ \\
    $\xi^{10}$  & $\omega_3\mathbbm{1}_3$ & $\text{Diag}(1,\omega_3,\omega_3^2)$ & $\chi$ \\
    $\xi^{11}$  & $\omega_3^2\mathbbm{1}_3$ & $\text{Diag}(1,\omega_3^2,\omega_3)$ & $\chi$ \\
    \end{tabular}
    \end{table}
    
    \begin{table}
\caption{The 9 irreps for $\Delta(54)$.}
\label{tab:su3irrepsd54}
    \begin{tabular}{c|cccc}
    g & $\omega_3$ & $\mathbf{C}$ & $\mathbf{E}$ & $\mathbf{V}^2$ \\\hline\hline
    $\bar{\rho}^1$ & 1 & 1 & 1 & 1\\
    $\bar{\rho}^2$ & 1 & 1 & 1 & $-1$\\
    $\bar{\rho}^3$ & $\mathbbm{1}_2$ & $\text{Diag}(\omega_3,\omega_3^2)$ & $\mathbbm{1}_2$ & $X$\\
    $\bar{\rho}^4$ & $\mathbbm{1}_2$ &$\mathbbm{1}_2$ & $\text{Diag}(\omega_3,\omega_3^2)$ &  $X$\\
    $\bar{\rho}^5$ & $\mathbbm{1}_2$ & $\text{Diag}(\omega_3^2,\omega_3)$ & $\text{Diag}(\omega_3,\omega_3^2)$ &  $X$\\
    $\bar{\rho}^6$ & $\mathbbm{1}_2$ & $\text{Diag}(\omega_3,\omega_3^2)$ & $\text{Diag}(\omega_3^2,\omega_3)$ &  $X$\\
     $\bar{\rho}^{7}$  & $\omega_3\mathbbm{1}_3$ & $\text{Diag}(1,\omega_3,\omega_3^2)$ & $\chi$ & $-X_{1,2}$\\
    $\bar{\rho}^{8}$  & $\omega_3^2\mathbbm{1}_3$ & $\text{Diag}(1,\omega_3^2,\omega_3)$ & $\chi$ & $-X_{1,2}$\\
    $\bar{\rho}^{9}$  & $\omega_3\mathbbm{1}_3$ & $\text{Diag}(1,\omega_3,\omega_3^2)$ & $\chi$ & $X_{1,2}$\\
    $\bar{\rho}^{10}$  & $\omega_3^2\mathbbm{1}_3$ & $\text{Diag}(1,\omega_3^2,\omega_3)$ & $\chi$ & $X_{1,2}$\\
    \end{tabular}
    \end{table}
    
    \begin{table}
        \caption{The 14 irreps, $\rho^a$ for $\Sigma(36\times3)$.}
        \label{tab:su3irrepss108}
    \begin{tabular}{c|ccccc}
    g & $\omega_3$ & $\mathbf{C}$ & $\mathbf{E}$ & $\mathbf{V}$\\\hline\hline
    $\rho^1$ & 1 & 1 & 1 & 1\\
    $\rho^2$ & 1 & 1 & 1 & $i$\\
    $\rho^3$ & 1 & 1 & 1 &$-1$\\
    $\rho^4$ & 1 & 1 & 1 & $-i$\\
    $\rho^5$ & $\omega_3\mathbbm{1}_3$ & $S_3$ & $\chi$ & -$iH_3$ \\
    $\rho^6$ & $\omega_3\mathbbm{1}_3$ & $S_3$ & $\chi$ & $H_3$\\
    $\rho^7$ & $\omega_3\mathbbm{1}_3$ & $S_3$ & $\chi$ & $iH_3$\\
    $\rho^8$ & $\omega_3\mathbbm{1}_3$ & $S_3$ & $\chi$ & $-H_3$\\
    $\rho^9$ & $\omega_3^2\mathbbm{1}_3$ & $S_3^\dag$ & $\chi$ & $-iH_3$\\
    $\rho^{10}$ & $\omega_3^2\mathbbm{1}_3$ & $S_3^\dag$ & $\chi$ & $H_3$\\
    $\rho^{11}$ & $\omega_3^2\mathbbm{1}_3$ & $S_3^\dag$ & $\chi$ & $iH_3$\\
    $\rho^{12}$ & $\omega_3^2\mathbbm{1}_3$ & $S_3^\dag$ & $\chi$ & $-H_3$\\
    $\rho^{13}$ & $\mathbbm{1}_4$ & $\text{Diag}(1,\omega_3,1,\omega_3^2)$ & $\text{Diag}(\omega_3,1,\omega_3^2,1)$ & $\chi_4$\\
    $\rho^{14}$ & $\mathbbm{1}_4$ & $\text{Diag}(\omega_3,\omega_3,\omega_3^2,\omega_3^2)$ & $\text{Diag}(\omega_3,\omega_3^2,\omega_3^2,\omega_3^2)$ & $\chi_4$
    \end{tabular}
\end{table}
\subsection{Fourier transformation for \texorpdfstring{$\Delta(27)$}{}}

We need to first identify the operator $FFT$ which generates the Fourier transformation for $\mathbb{Z}_3\times\mathbb{Z}_3$. 
This quantum circuit is simply the tensor product of a discrete abelian quantum Fourier transformation across each of the individual $\mathbb{Z}_3$ groups.
This leaves us with the irreps corresponding to the following basis vectors:
\begin{align}
\label{eq:z3z3basis}
    |00\rangle = &|\chi^1\rangle & |01\rangle = &|\chi^2\rangle\notag & |02\rangle = &|\chi^3\rangle\notag\\
    |10\rangle = &|\chi^4\rangle & |11\rangle = &|\chi^5\rangle&|12\rangle = &|\chi^6\rangle\notag\\
    |20\rangle = &|\chi^7\rangle & |21\rangle = &|\chi^8\rangle & |22\rangle = &|\chi^9\rangle
\end{align}

\begin{figure*}
\includegraphics[width=0.28\linewidth]{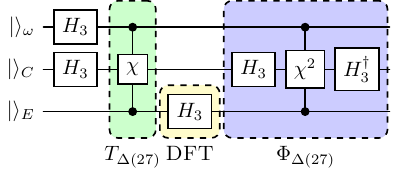}
\includegraphics[width=0.68\linewidth]{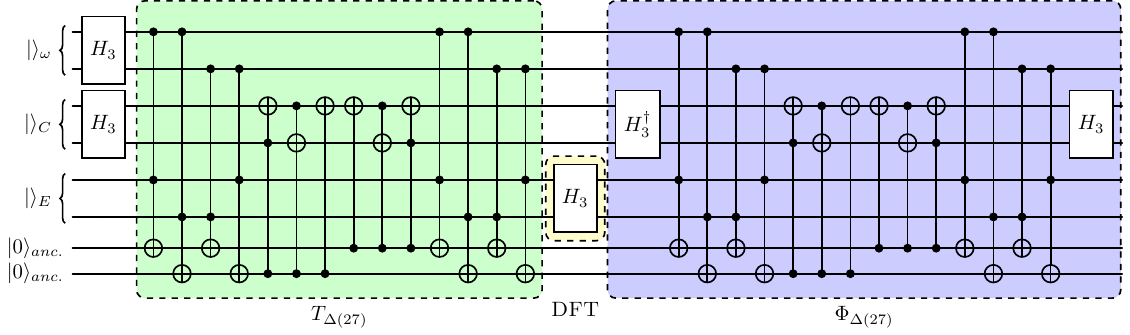}
\caption{Qutrit (left) and qubit (right) version of the $\mathfrak{U}_{FFT}^{\Delta(27)}$.}
\label{fig:dftd27qubit}
\end{figure*}

Now we need to identify which irreps form the conjugate pairs from \cite{Pueschel:1998zzo} and which are trivially extended.
This involves calculating what $\chi^{3a + b + 1}(\mathbf{E}^r \omega_3^p \mathbf{C}^q \mathbf{E}^{-r})$ is equal to. 
We find that $\lbrace \chi^{1},~\chi^2,~\chi^3\rbrace$ are trivial, i.e., $\chi^{m}(\mathbf{E}^r \omega_3^p \mathbf{C}^q \mathbf{E}^{-r}) = \chi^{m}$ for $m=1,2,3$, while $\lbrace \chi^4,~\chi^5,~\chi^6\rbrace$ and $\lbrace \chi^7,~\chi^8,~\chi^9\rbrace$ form inner conjugates.
We can now write down the operator for $T_{\Delta(27)}$,
\begin{equation}
    \label{eq:Td27}
    T_{\Delta(27)} = \sum_{r=0}^2 |r\rangle\langle r|\otimes \mathcal{T}_{\Delta(27)}^r
\end{equation}
where,
\begin{equation}
\label{eq:Dd27}
    \mathcal{T}_{\Delta(27)} = \mathbbm{1}_{3}\oplus\chi^{-1}\oplus \chi
\end{equation}
Additionally we can write down the phase kickback matrix,
\begin{align}
    \Phi_{\Delta(27)} = \sum_{r=0}^{r} |r\rangle\langle r| \otimes \phi_{\Delta(27)}^r
\end{align}
with
\begin{align}
\label{eq:Cd27}
    \phi_{\Delta(27)} = \mathbbm{1}_{3}\oplus\text{Diag}(1, \omega_3^2, \omega_3) \oplus \text{Diag}(1, \omega_3, \omega_3^2).
\end{align} 
These pieces in turn give us the necessary unitaries to construct the $\mathfrak{U}_{FFT}^{\Delta(27)}$. The operator in Eq.~(\ref{eq:Dd27}) is given by a inverted controlled sum gate, while the operator in Eq.~(\ref{eq:Cd27}) is a inverted controlled phase gate. We provide the full quantum circuit for qutrits in left panel of Fig.~\ref{fig:dftd27qubit} with each of the components highlighted and labeled. 
After this Fourier transformation we can identify the computational basis states with their corresponding irreps:
\begin{align}
    |000\rangle = & |\xi^1\rangle, & |001\rangle = & |\xi^2\rangle, & |002\rangle = & |\xi^3\rangle,\notag\\
    |010\rangle = & |\xi^4\rangle, & |011\rangle = & |\xi^5\rangle, & |012\rangle = & |\xi^6\rangle,\notag\\
    |020\rangle = & |\xi^7\rangle, & |021\rangle = & |\xi^8\rangle,& |022\rangle = & |\xi^9\rangle,\notag
\end{align}
\begin{align}
    \lbrace|1ab\rangle:~0\leq a,b\leq 2 \rbrace = & |\xi^{10}\rangle\notag\\
    \lbrace|2ab\rangle:~0\leq a,b \leq 2\rbrace = & |\xi^{11}\rangle 
\end{align}

While this qutrit decomposition for the fast Fourier transformation is the most straightforward to right down, it may not necessarily be the most efficient way as fault tolerant hardware is developed. In this context it is important to have a purely qubit based version of this fast Fourier transformation.

We provide the quantum circuit for $\mathfrak{U}_{FFT}^{\Delta(27)}$ using a homogenous qubit register in the right panel of Fig.~\ref{fig:dftd27qubit}. This circuit is the functional equivalent of the one provided in Fig.~\ref{fig:dftd27qubit}. Each of the three qutrits is broken down into a two qubit subregister. 

The main non-trivial component of the twiddle operation is how the multi-controlled shift operation is applied. We use the same rule as discussed in the previous section to map onto the first ancilla.
We compute the operators that correspond to a product equal to 2 modulo 3 on the first ancilla and a product equal to 1 modulo 3 on the second register. 
This reclaimable scratch register computation is done to minimizes the T-gate count for the quantum simulation. 
This operator ends up being repeated again for $\Phi_{\Delta(27)}$. 
In total this circuit requires 24 Toffoli operations.  
The remaining T-gates come from the decomposition of the $H_3$ gates discussed in the prior subsection. As mentioned above each of these operations requires 14 $R_z$ rotations and 3 transversal CNOTs. 
This gives us a total T-gate count of:
\begin{equation*}
168 + 80.5 \text{log}_2(1 / \epsilon),
\end{equation*}
where $\epsilon$ is the target synthesis error. The circuit has a maximal T-gate width of 4 and requires 2 ancilla; which corresponds to a temporary 33\% qubit memory overhead increase for the scratch registers.

\subsection{Fourier transformation for \texorpdfstring{$\Delta(54)$}{}}

As the derivation follows similarly to the previous ones, we will abbreviate it going further. We first need to identify the irreps of $\Delta(27)$ which form conjugate pairs.
We can identify this by observing how the operator $\mathbf{V}^2$ transfroms an arbitrary element of $\Delta(27)$; we find that
\begin{align}
    \mathbf{V}^2 \omega_3^p \mathbf{C}^q \mathbf{E}^r \mathbf{V}^2 = \omega_3^p \mathbf{C}^{2q}\mathbf{E}^{2r}.
\end{align}
This result implies that we have the following conjugate pairs from the $\Delta(27)$ irreps:
\begin{align}
    \{\xi^2,~\xi^3\},~\{\xi^4,~\xi^7\},~\{\xi^5,~\xi^9\},~\{\xi^6,~\xi^8\},~\{\xi^{10},~\xi^{11}\}\notag
\end{align}
These conjugate pairs allow us to identify the operator $\Phi_{\Delta(54)}$ which is given by
\begin{align}
\label{eq:d54pkb}
    \Phi_{\Delta(54)} = &\sum_{s=0}^1 |s\rangle\langle s| \otimes \notag\\\Bigg(
&\mathbbm{1}_2 \oplus -1 \oplus \mathbbm{1}_3 \oplus -\mathbbm{1}_3\oplus
    \mathbbm{1}_9\oplus-\mathbbm{1}_9\Bigg)^s.
\end{align}

\begin{figure*}
    \includegraphics[width=0.48\linewidth]{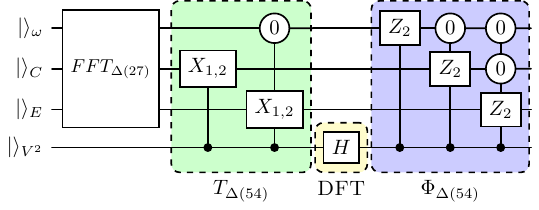}
        \includegraphics[width=0.48\linewidth]{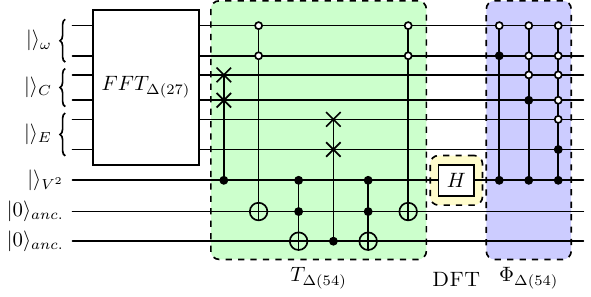}
    \caption{Quantum circuit of $\mathfrak{U}_{FFT}^{\Delta(54)}$ using a heterogenous qubit-qutrit register (left) and homogenous qubit register (right).}
    \label{fig:d54fft}
\end{figure*}

The twiddle matrix, $T_{\Delta(54)}$, is 
\begin{align}
    T_{\Delta(54)} = \sum_{s=0}^1 |s\rangle\langle s|\otimes \mathcal{T}_{\Delta(54)}
\end{align}
where
\begin{align}
    \mathcal{T}_{\Delta(54)} = &X_{1,2}\oplus(X\otimes X_{1,2}) \oplus\mathbbm{1}_{3}\oplus\notag\\
    &(X\otimes\mathbbm{1}_3)\oplus\mathbbm{1}_{3}\oplus(X\otimes\mathbbm{1}_3).
\end{align}
The first two terms of matrix sum in $\mathcal{T}_{\Delta(54)}$, correspond to the trivial irrep and the mixing of the other 1d irreps. The last terms in the sum are the mixing introduced by $\mathbf{V}^2$ to the 3d irreps $\xi^{10}$ and $\xi^{11}$.

The total fast Fourier transformation is provided in on the left panel of Fig. \ref{fig:d54fft}. In this figure both the twiddle and phase kickback operations highlighted as well as the Fourier transformation across the transversal. We now proceed to the details of the qubit decomposition for this fast Fourier transformation. Because the $\mathbf{V}^2$ register only requires a qubit, there is only the decomposition of the qutrit operations that are important. 

The decomposition to qubit gates is substantively simpler than in the target case for $\Delta(27)$. The controlled $X_{1,2}$ operations are mapped to controlled swaps while the $Z_2$ operations are mapped to controlled CZ gates. This decomposition using the results of the appendix provides us with the circuit in the right panel of Fig. \ref{fig:d54fft}. This circuit uses the same 2 ancilla as in the $\Delta(27)$ circuit. The decomposition requires 6 Toffoli gates for $T_{\Delta(54)}$ and 12 for $\Phi_{\Delta(54)}$ giving a total of 126 T gates on top of the $\Delta(27)$ costs. This yields a total T gate cost of 
\begin{equation}
    294 + 80.5 \text{log}_2(1 / \epsilon).
\end{equation}
This requires 5 ancilla qubits, the majority of which occur in the phase kickback operation. 

\subsection{Fourier transformation for \texorpdfstring{$\Sigma(36\times3)$}{}}
\begin{figure*}
    \includegraphics[width=0.4\linewidth]{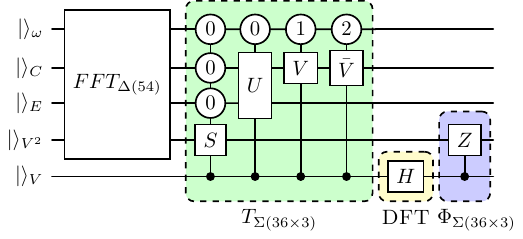}
        \includegraphics[width=0.58\linewidth]{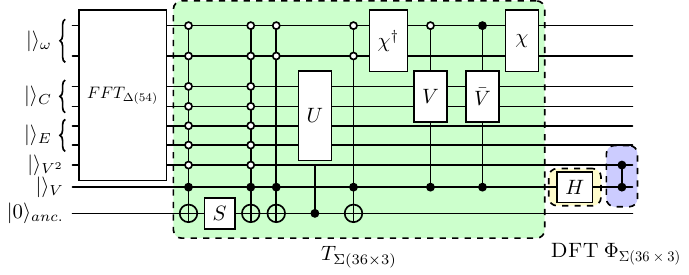}
    \caption{Quantum circuit implementing $\mathfrak{U}_{FFT}^{\Sigma(36\times3)}$ using a mixed qutrit-qubit encoding (left) and pure qubit register (right).}
    \label{fig:s108circuit}
\end{figure*}

\begin{figure}
 \includegraphics[width=\linewidth]{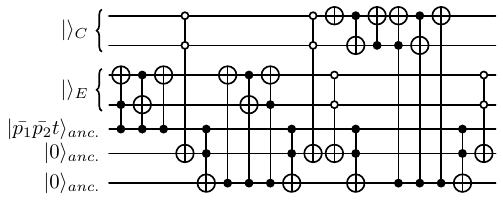}   \caption{Controlled $U$ operation from Fig. \ref{fig:s108circuit} using two clean ancilla and the precomputed value $\bar{p}_1\bar{p}_2t$.}
    \label{fig:Uqubit}

    \end{figure}

In order to progress from  $\Delta(54)$ to $\Sigma(36\times3)$, we need to construct a basis for the irreps of $\Delta(54)$.  We take the following elements:
\begin{align}
    |\bar{\rho}^1\rangle &=  |0000\rangle\qquad  |\bar{\rho}^2\rangle=|1000\rangle\notag\\
    |\bar{\rho}^3\rangle&=\lbrace |0100\rangle,|0200\rangle, |0210\rangle,|0020\rangle\rbrace \notag\\
    |\bar{\rho}^4\rangle&=\lbrace |1100\rangle,|1200\rangle, |1210\rangle,|1020\rangle\rbrace  \notag\\
    |\bar{\rho}^5\rangle &=\lbrace |0010\rangle,|0020\rangle,|0120\rangle,|0220\rangle\rbrace\notag\\
    |\bar{\rho}^6\rangle&=\lbrace |1000\rangle,|1110\rangle, |1120\rangle,|1220\rangle\rbrace \notag\\
    |\bar{\rho}^7\rangle&=\lbrace |0ab1\rangle:\,0\leq a,b \leq 2\rbrace\notag\\
    |\bar{\rho}^8\rangle&=\lbrace |1ab1\rangle:\,0\leq a,b \leq 2\rbrace \notag\\
    |\bar{\rho}^9\rangle&=\lbrace |0ab2\rangle:\, 0\leq a,b \leq 2\rbrace\notag\\
    |\bar{\rho}^{10}&=\rangle\lbrace |1ab2\rangle:\, 0\leq a,b \leq 2\rbrace.
\end{align}
We also need the following result from inner conjugation of $\Delta(54)$ by $\mathbf{V}$:
\begin{align}
    \mathbf{V} (\omega_3^p \mathbf{C}^q \mathbf{E}^r \mathbf{V}^{2s})\mathbf{V}^{-1} &= \omega_3^{p + 2qr} \mathbf{C}^{2r} \mathbf{E}^{q} \mathbf{V}^{2s}.
\end{align}
This tells us that the inner conjugates are $\lbrace \bar{\rho}^3,\bar{\rho}^4\rbrace$, $\lbrace \bar{\rho}^5,\bar{\rho}^6\rbrace$, $\lbrace \bar{\rho}^7,\bar{\rho}^8\rbrace$, $\lbrace \bar{\rho}^9,\bar{\rho}^{10}\rbrace$
This immediately allows us to identify that shuffling and transformation between the states needs to occur according to the representations in Tab. \ref{tab:su3irrepss108}. In total this gives us the twiddle operations
\begin{align}
    T_{\Sigma(36\times3)} = \sum_{t=0}^1 |t\rangle \langle t| \otimes \mathcal{T}_{\Sigma(36\times 3)}
\end{align}
with
\begin{align}
    &\mathcal{T}_{\Sigma(3\times 36)} = \begin{pmatrix}
        1 & 0\\
        0 & i\\
    \end{pmatrix}
    \oplus \Bigg(\mathbbm{1}_4 \otimes \chi_4\Bigg)\oplus \mathcal{V} \oplus (\mathcal{V}^*)
\end{align}
where $\chi_4$ is the 4d generalization of the clock-shift operation
\begin{equation}
    \chi_4=\begin{pmatrix}
        0 & 1 & 0 & 0\\
        0 & 0 & 1 & 0\\
        0 & 0 & 0 & 1\\
        1 & 0 & 0 & 0\\
    \end{pmatrix}
\end{equation}
and
\begin{align}
    \mathcal{V} = \mathbbm{1}_6\otimes (-iH_3).
\end{align}
Addtionally we can immediately identify the phase kickback operation as
\begin{align}
    \Phi_{\Sigma(36\times 3)} = \sum_{t} |t\rangle\langle t| \otimes \Bigg(\bigoplus_{j=0}^{27} \begin{pmatrix} 1 & 0\\0 & -1\end{pmatrix}\Bigg)^t.
\end{align}
The quantum circuit using a mixed qutrit-qubit system is provided in Fig.~\ref{fig:s108circuit}. The pure qubit decomposition into a fault tolerant cirucit is provided in the remainder of this subsection.
The $\Sigma(36\times3)$ fast Fourier transformation shown in Fig.~\ref{fig:s108circuit} has the decomposition to a 8 qubit register as shown in Fig.~\ref{fig:s108circuit}. 
Unlike the other two circuits where the twiddle matrix and phase kick back were approximately equal in T-gate costs. The twiddle matrix is the most expensive portion of the $\Sigma(36\times 3)$ fast Fourier transformation. The first controlled $S$ matrix requires 8 clean ancilla qubits for the scratch register. This scratch register bloat is only necessary at one small point for the controlled-$S$ gate. The explicit C$^n$NOTs that appear in the circuit require 20 Toffoli operations in total. There is an implicit set of 14 Toffoli gates within the controlled-$U$ operation which is shown explicitly in Fig. \ref{fig:Uqubit}.  

The final place where T-gate synthesis presents is in the controlled-$V$ and controlled-$\bar{V}$ case which follows from the derivation of Ref.~\cite{barenco1995elementary}. The first step is to diagonalize the $V$ operator, which will require 32 $R_z$ gates when dressing both sides of the diagonalizing operator. When diagonalized, the operator $V$ has in the qubit space, the diagonalized matrix,
\begin{align}
V_{diag.} = \text{Diag}(-i,i,1,1),
\end{align}
while $\bar{V}$ has the diagonalized form
\begin{align}
    \bar{V}_{\rm diag.} = \text{Diag}(i,-i,1,1).
\end{align}
These controlled operations together form a diagonal matrix which can be implementation with 4 $R_z$ and 6 CNOT gates using the methods of Ref.~\cite{PhysRevD.106.094504}. Therefore, in total the controlled $V$ and $\bar{V}$ incured a cost of 36 $R_z$ gates and no Toffoli gate. 
This all together brings the total number of additional T-gates on top of the $\Delta(54)$ gates to $238 + 36.8 \text{log}_2(1 / \epsilon)$ for a total T-gate count of
\begin{align}
    532 + 117.3 \text{log}_2(1 / \epsilon).
\end{align}

\begin{figure}
    \includegraphics[width=0.8\linewidth]{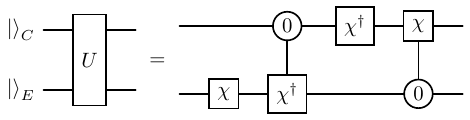}
    \caption{Quantum circuit of $U$ appearing in Fig.~\ref{fig:s108circuit}. $\chi$ is the qutrit increment gate, and $\chi^\dagger$ is the qutrit decrement gate. }
    \label{fig:us108}
\end{figure}

\section{Outlook for Quantum Simulations for QFTs}
\label{sec:outlook}
The results from this work allow us to make both concrete predictions for existing groups as well as extrapolations about the costs of simulations for various other nonabelian groups of import. The cost comparison for various decompositions are provided in Tab. \ref{tab:tgatecostbt}.

For one method of quantum simulation~\cite{PhysRevD.100.034518}, there are four primitive gates that must be constructed: the inverse, multiplication, trace and Fourier tranform gate;  which the Fourier transform dominating the resource costs. The main cost for quantum simulations involves mapping a total target synthesis error $\bar{\epsilon}_T^I$ to a total number of T-gates $N_{T}^{\rm{fid}}$. This comes from a target cost function $C_T^I$. Greater details about these methods and the finer nuances of these calculations can be found in Refs.~\cite{Kan:2021xfc, gustafson2024primitive}. 
The T-gate cost provided in this work can be considered quasi-optimal and an approximate lower bound for the total resources necessary for implementation.  For the groups $\mathbb{BT},\mathbb{BO},$ and $\Sigma{36\times3}$, the total simulation cost are presented in Tab.~\ref{tab:simcost}.

This reduction in cost for the $\mathbb{BT}$ Fourier transformation still leaves it as the most costly operation but nevertheless reduces the overall simulations costs for pure gauge viscosity simulations proposed in Ref.~\cite{Gustafson:2022xdt} by an additional order of magnitude and by 11 orders of magnitude compared to Ref.~\cite{Kan:2021xfc}. 
Meanwhile $\mathfrak{U}_{FFT}^{\mathbb{BO}}$ is nearly two orders of magnitude cheaper than $\mathfrak{U}_{FT}^{\mathbb{BO}}$. This brings the cost down for Fourier transformation to be on par with the trace and multiplication gates in terms of T-gate cost, bringing the total simulation cost down by 2 orders of magnitude. Even greater reductions are found with $\mathfrak{U}_{FFT}^{\Sigma(36\times3)}$ which reduces the cost of the gate by a factor of nearly three orders of magnitude at the expense of doubling the qubit overhead; resulting in a total reduction in simulation cost by a factor of 580. 
\begin{table}[t]
    \centering
    \caption{Number of physical T gates and clean ancilla required to implement logical gates for (first) basic gates taken from~\cite{Chuang:1996hw}, (second) primitive gates for $\mathbb{BT}$, (third) primitive gates for $\mathbb{BO}$, (fourth) primitive FFT for $\Delta(27)$, (fifth) primitive FFT for $\Delta(54)$, and (sixth) primitive gates for $\Sigma(36\times3)$.}
    \label{tab:tgatecostbt}
    \begin{tabular}{cccc}
    \hline\hline
         Gate & T gates & T gate width & Clean ancilla\\
         \hline
         C$^2$NOT & 7 & - & 0\\
         C$^3$NOT & 21 & 1 & 1\\
         $R_z$ & 1.15 $\log_2(1/\epsilon)$ & 1 & 0\\
         \hline
         $\mathfrak{U}^{\mathbb{BT}}_{FT}$ \footnote{\label{refbt}from \cite{Gustafson:2022xdt}} & 3735.2 $\log_2(1/\epsilon)$ & 5 & 0\\
        $\mathfrak{U}^{\mathbb{BT}}_{FT}$ \footnote{\label{refbo}from \cite{Gustafson:2023kvd}} & 2802.55 $\log_2(1/\epsilon)$ & 5 & 0\\
         $\mathfrak{U}^{\mathbb{BT}}_{FFT}$ & 98+48.3 $\log_2(1/\epsilon)$ & 2 & 2\\\hline
         $\mathfrak{U}^{\mathbb{BO}}_{FT}$ \footref{refbo}& 11370.1 $\log_2(1/\epsilon)$ & 6 & 0\\
         $\mathfrak{U}^{\mathbb{BO}}_{FFT}$ & 216+48.3 $\log_2(1/\epsilon)$ & 2 & 4\\\hline
    $\mathfrak{U}^{\Delta(27)}_{FFT}$ & 168 + 80.5 log$_2(1/\epsilon)$ & 4 & 2\\\hline
    $\mathfrak{U}^{\Delta(54)}_{FFT}$ & 294 + 80.5 log$_2(1/\epsilon)$ & 4 & 5\\\hline
    $\mathfrak{U}^{\Sigma(36\times3)}_{FT}$ \footnote{\label{refs108}from \cite{gustafson2024primitive}} &$185898 \log_2(1/\epsilon)$ & 4 & 0\\
    $\mathfrak{U}^{\Sigma(36\times3)}_{FFT}$ & $532 + 117.3 \text{log}_2(1 / \epsilon)$ & 8 & 8\\\hline\hline
    \end{tabular}
\end{table}

\begin{table*}
    \caption{T-gate count for a fiducial quantum simulation $N_{T}^{fid}$ and the factor of improvement from using our FFT $\mathcal{R}_{QFT}$.}
    \label{tab:simcost}
    \centering
    \begin{tabular}{c|c|c|c|c|c}
        $G$ &QFT Impl.& $C^I_T$  & $\tilde{\epsilon}^I_T$ & $N^{\rm fid}_T$&$\mathcal{R}_{QFT}$\\\hline\hline
         \multirow{2}{4em}{\centering$\mathbb{BT}$}& FT& $4676 d-3948 + (11191.2 + 18.975 d)\log_2\frac{1}{\epsilon}$  & $\frac{1}{2}(19463 + 33  d)$ &$9.8\times 10^{10}$ &\multirow{2}{4em}{\centering 29 } \\
         &FFT& $4676 d -3556 + (174.225 + 18.975d)\log_2\frac{1}{\epsilon}$ & $\frac{3}{2} (101 + 11  d)$ &$3.4\times 10^{9}$ &\\\hline
         \multirow{2}{4em}{\centering$\mathbb{BO}$}&FT& $11949d-10157+(45473.3+ 6.9d)\log_2\frac{1}{\epsilon}$ & $2 (19771 + 3 d)$ &$4.1\times10^{11}$ &\multirow{2}{4em}{\centering 73} \\
         & FFT & $11949d-9293 + (186.3 + 6.9  d)\log_2\frac{1}{\epsilon}$ &$6  (27 + d)$ &$5.6\times10^{9}$&\\\hline
         \multirow{2}{4em}{$\Sigma(36\times3)$} &FT&$9632d-8192+(744167+12.075d)\log_2\frac{1}{\epsilon}$& $\frac{3}{2}(431401+7d)$ &$7.0\times10^{12}$ &\multirow{2}{4em}{\centering 580}\\
         &FFT&$9632d-6034 + (1045.93 + 12.075  d)\log_2\frac{1}{\epsilon} $  & $\frac{1}{2} (1819 + 21 d)$ &$1.2\times10^{10}$ &\\\hline
    \end{tabular}
\end{table*}

\section{Discussion}
\label{sec:dis}
In this work, we have derived fast quantum Fourier transforms for a number of nonabelian groups of interest in high energy physics and beyond. A potentially interesting next step relevant to a wide variety of systems would be the exploring applications and tradeoffs involving qudits or mixed qubit and qudit systems. For example the symmetric and alternating groups, $\mathcal{S}_n$ and $\mathcal{A}_n$, accelerate in size quite quickly with the number of elements being $n!$ and $n!/2$ where $n$ is the number of elements being ``permuted". This factorial scaling indicates certain qubit memory may be wasted. For example with the symmetric group of the amount of ``wasted" states can be between 20-45\% for $n \leq 20$. This might demonstrate that there are hardware advantages to the use of qudit based systems for these sorts of problems.

The general structure of these fast Fourier transformations for smaller groups is paramount for the construction of fast Fourier transformations for larger ones. Hence results of this paper are applicable to further (super)groups in the same way. In particular the results of $\Sigma(36\times3)$ will be crucially important for the groups, $\Sigma(72\times3)$, $\Sigma(216\times3)$, and $\Sigma(360\times3)$, which are crucial for simulations of quantum chromodynamics. Additionally, it is important to note that our formulation of the quantum Fourier transform relied on a particular encoding of the discrete groups, and therefore modifications may be required for other encodings~\cite{Lamm:2024jnl}.  For quantum simulations of lattice gauge theories which use other digitizations, the implementation of fast Fourier transforms requires more investigation.

It worth also noting that the notions of approximations to the standard $\mathbb{Z}_n$ quantum Fourier transformation exist \cite{barenco1996approximate, Coppersmith:2002skh, hales2002quantum,  cheung2003using, cheung2004improved, rotteler2008representation, prokopenya2015approximate, nam2020approximate}. An fascinating direction is to develop analogous constructions for so-called approximate quantum Fourier transformations for nonabelian groups, as well as identify potential applications where such approximate implementations yield potential advantages over exact implementations. 

One may wish to further explore applications of novel insights related to nonabelian QFTs to the HSP. Various papers have considered quantum algorithms for the nonabelian HSP, see Refs.~\cite{ivanyos2001efficient,regev2004subexponential,bacon2005optimal,radhakrishnan2005power,ivanyos2007efficient,kuperberg2011another},  and this work has demonstrated explicitly how to constructions the fast Fourier transformation for nontrivial example groups; a crucial step forward toward efficient algorithms for HSP. The research performed in this work demonstrates that nonabelian fast Fourier transformations can be constructed efficiently for fault tolerant qubit based hardware and can offer orders-of-magnitude reduction in T-gate cost for quantum simulation. 
\begin{acknowledgements}
    The authors would like to thank Namit Anand, Aaron Lott, and Norman Tubman for helpful comments and suggestions. We are grateful for support from NASA Ames Research Center. Erik Gustafson, Stuart Hadfield, and M. Sohaib Alam were supported by the NASA Academic Mission Services, Contract No. NNA16BD14C and NASA-DOE interagency agreement SAA2-403602, and through the Superconducting Quantum and Materials System Center (SQMS) under contract number DE-AC02-07CH11359. This material is based on work supported by the U.S. Department of Energy, Office of Science, National Quantum Information Science Research Centers, Superconducting Quantum Materials and Systems Center 
    under contract number DE-AC02-07CH11359 (Henry Lamm). Edison Murairi is supported under the grant number DE-FG02-95ER40907. Fermilab is operated by Fermi Research Alliance, LLC under contract number DE-AC02-07CH11359 with the United States Department of Energy.
\end{acknowledgements}

\bibliography{ref,refs2}

\end{document}